\documentclass[12pt, a4paper]{article}
\pdfoutput=1

\usepackage[height=22cm,width=16cm, centering]{geometry}
\usepackage{setspace}

\usepackage[utf8]{inputenc}

\usepackage{amsmath}
\usepackage{amssymb, bbold, bm}
\usepackage{graphicx}
\usepackage{color,comment, slashed}
\usepackage[sort&compress, numbers, merge]{natbib}

\definecolor{blue-purple}{rgb}{0.25 ,0, 0.75}
\definecolor{red-purple}{rgb}{0.5 ,0, 0.6}
\definecolor{bluegreen}{rgb}{0, 0.35, 0.45}
\usepackage{hyperref}
\hypersetup{colorlinks=true, linkcolor=bluegreen, citecolor=blue-purple, urlcolor=red-purple}

\def\Mpl{M_{\rm P}}
\def \psiL{\psi^{\ell}}
\def \psiLc{\psiL_{\text{c}}}
\def \psiT{\psi^t}
\def \vecpsiT{\vec{\psi}^{t}}

\allowdisplaybreaks

\begin{document}

\begin{titlepage}
\begin{center}
\leavevmode \\
\vspace{ 0cm}

\hfill {\small CTPU-PTC-21-15}

\noindent
\vskip 1 cm
\textbf{\fontsize{15.8pt}{0pt} \selectfont Minimal Supergravity Inflation without Slow Gravitino}

\vglue .5in

{\fontsize{14pt}{0pt} \selectfont 
 Takahiro Terada
}

\vglue.3in

\textit{
Center for Theoretical Physics of the Universe, \\ Institute for Basic Science (IBS),
  Daejeon 34126, Korea
}
\end{center}

\vglue 0.2in

\begin{abstract}
We utilize a recently proposed cubic nilpotent superfield to realize inflation in supergravity with the minimal degrees of freedom: the inflaton, graviton, and massive gravitino.  As an advantage, the resultant model is free from the catastrophic production of gravitinos due to its vanishing propagation speed.  However, the model suffers from the standard gravitino problem, and its viability depends on the mass spectrum and the thermal history of the universe. 
\end{abstract}
\vglue 6mm

\tableofcontents

\end{titlepage}

\section{Introduction \label{sec:intro}}

If the fundamental theory has supersymmetry, its cosmological application involves supergravity and its spontaneous breaking.  Supersymmetry breaking in cosmology can be triggered by any positive energy density such as the kinetic and/or potential energy of a scalar field and a thermal environment.  In particular,  a nonzero Hubble parameter $H$ always contributes to supersymmetry breaking.  Since this breaking is of the same order as the cosmologically relevant scale $H$, it is reasonable to use the spontaneously broken supersymmetric theory rather than completely general non-supersymmetric theories.  

Speaking about the observational side, the cosmological data are consistent with single-field slow-roll inflation since there are no significant isocurvature perturbations~\cite{Akrami:2018odb} and non-Gaussianity~\cite{Akrami:2019izv}.  This implies that other fields than inflaton are sufficiently heavy and they are (semi)decoupled during inflation. In supergravity, generic fields are expected to have the Hubble-induced mass of $\mathcal{O}(H)$, opening up an interesting window for future observations~\cite{Chen:2009we, Chen:2009zp, Baumann:2011nk, Noumi:2012vr, Arkani-Hamed:2015bza, Chen:2016uwp, Arkani-Hamed:2018kmz}.

 In particular, the scalar superpartner of the inflaton (sinflaton) should also become heavy.  Typical (inflation) model building gives sinflaton (and/or a stabilizer field~\cite{Kawasaki:2000yn, Kallosh:2010ug, Kallosh:2010xz}) a mass much larger than $\mathcal{O}(H)$ for its strong stabilization~\cite{Dine:2006ii, Kitano:2006wz, Kallosh:2006dv, Lee:2010hj, Evans:2013nka, Ellis:2013nxa} and it stays heavier than the Hubble scale also after inflation.  When a part of the supermultiplet is decoupled, supersymmetry is non-linearly realized~\cite{Volkov:1972jx, Volkov:1973ix}. Constrained superfields  like a nilpotent chiral superfield $X$ satisfying $X^2 = 0$ are useful tools to describe such a system~\cite{Rocek:1978nb, Ivanov:1978mx, Lindstrom:1979kq, Casalbuoni:1988xh, Komargodski:2009rz}.\footnote{
See also Refs.~\cite{Kuzenko:2010ef, DallAgata:2016syy, Cribiori:2017ngp} for theoretical developments. For cosmological applications of nilpotent superfields, see Refs.~\cite{Antoniadis:2014oya, Ferrara:2014kva, Kallosh:2014via, Kallosh:2014hxa, DallAgata:2014qsj, Dudas:2016eej, Argurio:2017joe, Dalianis:2017okk}.  The inflaton potential is usually constructed from the superpotential, but it can also arise from the coupling between the inflaton and the nilpotent field in the K\"ahler potential~\cite{McDonough:2016der, Kallosh:2017wnt}.  However, one needs to be careful about the gravitino problem due to the same coupling~\cite{Hasegawa:2017nks}.
}
For example, if the sinflaton is in the lowest component of $T + \bar{T}$ with $T$ being a chiral superfield, it can be eliminated from the spectrum by the constraint $\bar{X} X (T + \bar{T}) = 0$~\cite{DallAgata:2016syy}.

It is then natural to ask whether the fermionic superpartner, inflatino, can also be eliminated from the inflaton supermultiplet.  If this is indeed the case, it has a significant impact on cosmology.  On one hand, we do not need to worry about the cosmological inflatino problem~\cite{Nilles:2001my}.  On the other hand, the analysis of the gravitino production after inflation can be significantly simplified given that the diagonalization between the gravitino-inflatino system~\cite{Nilles:2001ry, Nilles:2001fg, Ema:2016oxl, Roberts:2021plm} is no longer needed.  Such a setup involves the minimal degrees of freedom for inflation in supergravity: the (real) inflaton, gravitons, and massive gravitino.  It has been studied in terms of the component fields~\cite{Delacretaz:2016nhw} (see also Ref.~\cite{Cribiori:2016qif}) or in terms of constrained superfields called the orthogonal nilpotent superfields~\cite{ Kahn:2015mla, Ferrara:2015tyn, Carrasco:2015iij} $X$ and $T$ satisfying $X^2 = X(T+\bar{T})= 0$, which also implies $(T+\bar{T})^3 = 0$~\cite{Komargodski:2009rz}.  The models in Refs.~\cite{Ferrara:2015tyn, Carrasco:2015iij} were dubbed as the ``minimal supergravity inflation.'' 

The dynamics of the minimal supergravity inflation~\cite{Ferrara:2015tyn, Carrasco:2015iij} after inflation has been studied in Ref.~\cite{Hasegawa:2017hgd}.  It turned out that an anomalously significant gravitino production happens in this model due to a change of the propagation speed (also called ``sound speed'') of the longitudinal mode of the gravitino.  This phenomenon does not look like any other significant particle production we encounter,  e.g., in preheating~\cite{ Kofman:1997yn, Amin:2014eta} because the production in some limit does not depend on the wavenumber of the mode, so any modes below the cutoff scale, whatever the cutoff scale is, are produced.  Roughly speaking, this happens when the sound speed vanishes (see Ref.~\cite{Hasegawa:2017hgd} for more precise conditions and discussion on the backreaction).  If the theory is theoretically sound, the gravitino production is phenomenologically catastrophic, but this strange behavior lets one suspect that  the theory might be pathological.  This issue has recently been revisited in Refs.~\cite{Kolb:2021xfn, Kolb:2021nob}, including the case of the non-supersymmetric Rarita-Schwinger field, and the authors proposed a condition that forbids the catastrophic production of the ``slow gravitinos'' as a Swampland conjecture~\cite{Vafa:2005ui, Ooguri:2006in, Brennan:2017rbf, Palti:2019pca}.  Potential causes of the problem in a UV theory were recently pointed out in Ref.~\cite{Dudas:2021njv}.

Recently, Aldabergenov, Chatrabhuti, and Isono proposed an alternative constrained superfield~\cite{Aldabergenov:2021obf} that has the same independent degrees of freedom with the orthogonal nilpotent superfields.\footnote{
A version of the proposed constraints already appeared in appendix E of Ref.~\cite{Komargodski:2009rz}. 
A cubic nilpotent constraint on a deformed real linear superfield was proposed in Ref.~\cite{Kuzenko:2017oni}, which also leaves us the minimal degrees of freedom. See Ref.~\cite{Aldabergenov:2021obf} for more details on the relation between these approaches.} The same degrees of freedom do not necessarily imply the same interactions.  An important difference between these constraints is that the new constraint~\cite{Aldabergenov:2021obf} is imposed on a single chiral superfield whereas the orthogonal nilpotency conditions~\cite{Komargodski:2009rz, Kahn:2015mla, Ferrara:2015tyn, Carrasco:2015iij, DallAgata:2015zxp} are imposed on two chiral superfields.  As we will see, this difference is crucial in determining the properties of the longitudinal gravitino.

In this paper, we realize inflation in supergravity with minimal degrees of freedom without the problem of the catastrophic gravitino production due to the change of the sound speed by utilizing the new constrained superfield of Ref.~\cite{Aldabergenov:2021obf}.  
In section~\ref{sec:constraint}, we review the constraint proposed in Ref.~\cite{Aldabergenov:2021obf} and derive the solution of the constraint in the unitary gauge of supergravity.  Section~\ref{sec:inflation} describes inflation in our setup.  The dispersion relation and the production of gravitinos are discussed in section~\ref{sec:gravitino}.  We will see that there is no catastrophic gravitino production due to the sound-speed change, but we also discuss the ``standard'' gravitino problem~\cite{Pagels:1981ke, Weinberg:1982zq, Khlopov:1984pf,Ellis:1984eq} which is not directly related to the sound speed of the gravitino.   Section~\ref{sec:discussion} contains a summary and discussions. 

 For simplicity of presentation, we assume real parameters in the main text, and the general case is studied in three appendices.  In appendix~\ref{sec:model}, we introduce another example of an inflation model that has complex parameters.  The gravitino Lagrangian and the equations of motion are studied for the single-superfield case in the presence of the time-dependent phase of the gravitino mass and the nonvanishing vector auxiliary field in supergravity.  This appendix applies also to the standard supergravity without constrained superfields.  We summarize the general dispersion relation for a fermion with $\gamma^0$ and $\gamma_* \equiv i \gamma_0 \gamma_1 \gamma_2 \gamma_3$ dependence in its Lagrangian in appendix~\ref{sec:dispersion}. 
We use the mostly plus sign convention for the metric and the reduced Planck unit $c=\hbar = M_\text{P} (\equiv (8\pi G)^{-1/2}) = 1$, except when an explicit appearance of $M_\text{P}$ may help.

\section{Cubic nilpotent supermultiplet and the unitary gauge in supergravity \label{sec:constraint}}
Recently, a new cubic constraint on a supermultiplet was proposed in Ref.~\cite{Aldabergenov:2021obf}.
This can be viewed as a way to describe a low-energy effective field theory (EFT) of a spontaneously broken Abelian symmetry.
Depending on the linear or nonlinear representation of the (pseudo-)Nambu-Goldstone mode, they proposed two versions of the constraint.
The constraint on a chiral superfield $\Phi$ in the non-linear (shift-symmetric) version is 
\begin{align}
(\Phi+ \bar{\Phi})^3 = 0, \label{cubic_constraint_shift}
\end{align}
where the possible vacuum expectation value (VEV) of $\Phi$ has been subtracted for simplicity. (More generally, the constraint reads $\Sigma^3 = 0$ with $\Sigma \equiv \Phi + \bar{\Phi} - \langle \Phi + \bar{\Phi} \rangle$.)  The constraint on a chiral superfield $Z$ in the linear (U(1)-symmetric) version is
 $(Z \bar{Z} - \langle Z \bar{Z} \rangle )^3= 0$.  
In the following, we consider the former type [eq.~\eqref{cubic_constraint_shift}], since it may also describe the EFT of not only an axion-like field but also a dilaton-like field. In that case, we consider a constraint $(\Phi - \bar{\Phi})^3 = 0$.\footnote{
At this point, these are equivalent via a holomorphic field redefinition $\Phi \to i \Phi$.  The differences are a matter of convention.  In general, however, the axionic direction and the dilatonic direction are physically inequivalent since the former has a periodicity (an exact, non-perturbative, and discrete shift symmetry).  We use the word ``dilatonic'' here in a loose sense to mean that it does not have the exact shift symmetry. We introduce this distinction because example inflation models in the following sections do not have periodicity. 
}  We call $\Phi$ a cubic nilpotent superfield.

The constraint~\eqref{cubic_constraint_shift} is solved at the global supersymmetry level in Ref.~\cite{Aldabergenov:2021obf}.  We quote the result for $\Sigma \equiv \Phi + \bar{\Phi}$,  
\begin{align}
\Sigma = \chi^2 \beta + \bar{\chi}^2 \bar{\beta} + \frac{2}{U} \chi \sigma^\mu \bar{\chi} \partial_\mu \varphi, \label{constraint_solution}
\end{align}
where $\varphi = \text{Im} \, \Phi$ is the imaginary part of the lowest component of $\Phi$, $\chi$ is the Weyl-spinor component of $\Phi$, $U \equiv 2(|F|^2 - \partial_\mu \varphi \partial^\mu \varphi)$, $F$ is the auxiliary $F$-component of $\Phi$, and $\beta$ is given by~\cite{Aldabergenov:2021obf}
\begin{align}
\beta \equiv & \frac{\bar{F}}{U} + \frac{i \bar{\chi}}{U^2} \left( \bar{F} \bar{\sigma}^\mu \partial_\mu \chi - \partial_\mu \varphi\partial^\mu \bar{\chi} + 2 \partial_\mu \varphi \bar{\sigma}^{\mu\nu} \partial_\nu \chi \right) 
 - \frac{\bar{\chi}^2}{2 U^3} \left( F\partial_\mu \bar{\chi} \bar{\sigma}^{\mu\nu} \partial_\nu \bar{\chi} + \bar{F} \partial_\mu \chi \sigma^{\mu\nu} \partial_\nu \chi  \right. \nonumber \\
 & \qquad \qquad  \left. \phantom{  \bar{F} \partial_\mu \chi \sigma^{\mu\nu} \partial_\nu \chi } + \partial_\mu \varphi \partial_\nu \chi \left( 2\sigma^\mu \eta^{\nu\rho} - \sigma^\nu \eta^{\rho \mu} - \sigma^\rho \eta^{\mu\nu} - i \epsilon^{\mu\nu\rho\sigma} \sigma_\sigma \right) \partial_\rho \bar{\chi} \right) \label{beta}
\end{align}
in the two-component spinor notation~\cite{Wess:1992cp}.  
This shows that the scalar component of $\Sigma \equiv \Phi + \bar{\Phi}$ is not an independent dynamical scalar field but a composite field made of fermion bilinear terms and higher-order terms in fermions.  The solution will become more complicated when we consider supergravity; the partial derivative will become a covariant derivative of supergravity, which includes the gravitino field among others. 

Though we are interested in cosmological applications, we are not aiming at obtaining a general solution to the cubic nilpotent constraint in general gauges in supergravity.
A particularly useful gauge is the unitary gauge.
The (would-be) Nambu-Goldstone fermion (Goldstino) $\upsilon$ in supergravity is defined by~\cite{Freedman:2012zz}
\begin{align}
\upsilon_\text{L} \equiv e^{K/2} D_i W \chi_\mathrm{L}^i + g_{i \bar{j}} \partial_\mu  \phi^i \gamma^\mu \chi_\text{R}^{\bar{j}}, \label{goldstino}
\end{align}
where the subscripts L and R denote the left or right handedness, the sub/superscript $i$ and its conjugate $\bar{i}$ denote (super)field species, $K$ and $W$ are the K\"ahler potential and the superpotential, $g_{i\bar{j}}\equiv K_{i \bar{j}} $ is the K\"ahler metric, and  $D_i W \equiv W_i + K_i W$ is the K\"ahler covariant derivative.  Field derivatives are denoted by subscripts, e.g., $W_i \equiv \partial W/ \partial \phi^i$.  We do not consider the extension to include gauge fields since it is not essential in our discussion.

In the next section, we consider inflation models with a single chiral superfield $\Phi$ that obeys the cubic nilpotent constraint.
Ultimately, other fields are to be introduced to describe, e.g., the Standard Model fields, but the idea is that the inflation dynamics itself is dominated by the single chiral supermultiplet $\Phi$.  We assume therefore that no other fields contribute to the Goldstino. In particular, the relevant sum of the indices $i$ and $\bar{j}$ in the definition of the Goldstino only involves the inflaton supermultiplet. Note that the potential energy of the inflaton breaks supersymmetry through its nonzero $F$-term and the kinetic energy of the inflaton also breaks supersymmetry, which becomes relevant especially after inflation.  These correspond to the first and second term in eq.~\eqref{goldstino}, respectively.   The absence of other supermultiplets in the Goldstino implies the simple unitary gauge condition as follows: 
\begin{align}
\upsilon = \chi = 0 \qquad (\text{unitary gauge}), \label{unitary_gauge}
\end{align}
where $\chi$ is the fermionic component of $\Phi$, which is identified as the inflatino.

In the following, we exploit the unitary gauge condition to obtain the solution of the cubic nilpotent constraint~\eqref{cubic_constraint_shift} in supergravity.
In the tensor-calculus~\cite{Kaku:1978nz, Kaku:1978ea, Kugo:1983mv} notation, the components of the general supermultiplet are packaged as $(C, \zeta, H, K, v_\mu, \lambda, D)$~\cite{Stelle:1978yr, Kugo:1982cu}.  The chiral supermultiplet $\Phi$ in this notation is~\cite{Ferrara:2015tyn}
\begin{align}
\Phi = \left( \phi + i \varphi, - i \chi_\text{L}, - i F, - F, - i \hat{D}_\mu (\phi + i \varphi), 0, 0 \right),
\end{align}
where $\hat{D}$ is the supercovariant derivative ($\hat{D}_\mu \Phi = \partial_\mu \Phi - \frac{i}{2} \psi_\mu \chi_\mathrm{L}$ where $\psi_\mu$ is the gravitino; see Ref.~\cite{Ferrara:2015tyn} and references therein for the action on other fields).
In the unitary gauge with $\chi = 0$, this simplifies to 
\begin{align}
\Phi = \left( \phi + i \varphi, 0, - i F, - F, - i \partial_\mu (\phi + i \varphi), 0, 0 \right).
\end{align}
Using the multiplication rule~\cite{Stelle:1978yr, Kugo:1982cu}, we obtain 
\begin{align}
\frac{1}{8} ( \Phi + \bar{\Phi})^3 = \left( \phi^3, 0, 3 \phi^2 \text{Im}\,F , -3 \phi^2 \text{Re}\, F, 3 \phi^2 \partial_\mu \varphi , 0, 3\phi \left( -\partial^\mu \phi\partial_\mu \phi - \partial^\mu \varphi \partial_\mu \varphi + |F|^2 \right) \right).
\end{align}
In the unitary gauge, the lowest component $\phi^3 = 0$ implies $\phi = 0$.  This in turn implies that all the components of $(\Phi+\bar{\Phi})^3$ vanish and the constraint $(\Phi+\bar{\Phi})^3 = 0$ is satisfied.  Note that the auxiliary $F$-component is not constrained. 
We conclude that $\phi = \chi = \upsilon = 0$ in the unitary gauge, and the only remaining fields are the ``axion'' $\varphi$ and the auxiliary field $F$.
If we consider $(\Phi-\bar{\Phi})^3 = 0$ instead, then we obtain $\varphi = \chi = \upsilon = 0$ with the ``dilaton'' $\phi$ dynamical.

\section{Inflation with minimal degrees of freedom in supergravity \label{sec:inflation}}

In this section, we consider an inflation model utilizing the cubic nilpotent constraint.
This allows the minimal degrees of freedom to describe inflation in supergravity: the inflaton, graviton, and massive gravitino. 
There is no sinflaton.  Since there is no other candidate for Goldstino than inflatino, it is absorbed by the gravitino and eliminated from the spectrum. 
Inflation in supergravity with such minimal degrees of freedom has been studied in Refs.~\cite{Kahn:2015mla, Ferrara:2015tyn, Carrasco:2015iij, Delacretaz:2016nhw}.
We consider an alternative realization of this scenario utilizing a chiral superfield $\Phi$ with the constraint
\begin{align}
(\Phi - \bar{\Phi})^3 = 0. \label{inflaton_constraint}
\end{align}
 As we saw in section~\ref{sec:constraint}, the sinflaton and the inflatino vanish in the unitary gauge, but the $F$-term is not eliminated by the constraint. 
Without the constraint, the models we consider fit in the frameworks of the so-called sGoldstino inflation or the single-superfield inflation~\cite{Goncharov:1983mw, Izawa:2007qa, Achucarro:2012hg, AlvarezGaume:2010rt, AlvarezGaume:2011xv, Ketov:2014qha,Ketov:2014hya,Linde:2014hfa, Roest:2015qya, Linde:2015uga, Scalisi:2015qga, Ketov:2016gej, Ferrara:2016vzg}.

The general K\"ahler potential is expanded up to the second order in $(\Phi - \bar{\Phi})$ because of the constraint $(\Phi - \bar{\Phi})^3 = 0$,
\begin{align}
K (\Phi, \bar{\Phi}) =  K_0 (\Phi + \bar{\Phi} )  - i K_1 (\Phi + \bar{\Phi})  ( \Phi - \bar{\Phi}) - \frac{1}{2} K_2 (\Phi + \bar{\Phi}) ( \Phi - \bar{\Phi})^2 ,
\end{align}
where $K_0 (\Phi + \bar{\Phi} )$, $K_1 (\Phi + \bar{\Phi} )$, and $K_2 (\Phi + \bar{\Phi} )$ are real functions of $\Phi + \bar{\Phi} $.
 We assume the shift symmetry in the K\"ahler potential to avoid the $\eta$ problem~\cite{Kawasaki:2000yn}, so that $K_0$, $K_1$, and $K_2$ become constant. 
Since $K_0$ is simply absorbed by the redefinition of the superpotential by a constant factor\footnote{
Since the linear term in $K$ is holomorphic, it can also be absorbed by the superpotential with the replacement $W(\Phi) \to e^{-ic\Phi} W(\Phi)$.  The two frames are equivalent up to quantum anomaly~\cite{Bagger:2000dh}. 
} and $K_2$ just determines the normalization of the field, we consider a generic holomorphic superpotential $W(\Phi)$ and the following K\"ahler potential: 
\begin{align}
K(\Phi, \bar{\Phi}) = - i c (\Phi - \bar{\Phi}) - \frac{1}{2} (\Phi - \bar{\Phi})^2, \label{model_Kahler}
\end{align}
 where $c$ is a real constant.  Note that the superpotential breaks the shift symmetry, and this is understood as a soft breaking since the magnitude of $W$ is much smaller than the Planck scale to fit the cosmological data.

The scalar potential in the unitary gauge is given by the standard formula in supergravity 
\begin{align}
V =& e^K \left( g^{\bar{\Phi}\Phi} |D_\Phi W|^2 - 3 |W|^2 \right) \nonumber \\
=& |W_\Phi|^2 + \left(c^2 - 3\right) |W|^2 + 2 c \, \text{Im} \left(W \overline{W}_{\bar{\Phi}} \right), \label{V_general}
\end{align}
where $g^{\bar{\Phi} \Phi}$ is the inverse K\"ahler metric,  because the auxiliary component of $\Phi$ is not removed by the constraint.  This is in contrast to the setup in Refs.~\cite{Kahn:2015mla, Ferrara:2015tyn, Carrasco:2015iij}. 
 The last term vanishes if we assume a real holomorphic superpotential, i.e., $\overline{W}(\Phi) = W(\Phi)$~\cite{Kallosh:2010xz}. 
Note that we have used the fact that the sinflaton ($\text{Im}\, \Phi$) vanishes in the unitary gauge. 
The above formulas resemble those in Refs.~\cite{Ketov:2014qha,Ketov:2014hya, Ketov:2016gej}.  If we use the phase-symmetric constraint $(\Phi \bar{\Phi} - \langle \Phi \bar{\Phi} \rangle)^3 = 0$ instead of eq.~\eqref{inflaton_constraint}, we obtain formulas similar to those in Ref.~\cite{Ketov:2015tpa}.  On the other hand, the mathematical construction in Ref.~\cite{Ferrara:2016vzg} allows us to build any inflaton potential with a tunable cosmological constant and supersymmetry breaking (and hence the gravitino mass) at the vacuum.  

We classify two possibilities:
\begin{enumerate}
\item The inflaton supermultiplet breaks the supersymmetry at the vacuum, $\langle D_\Phi W \rangle \neq 0$ [$U\neq 0$ in eq.~\eqref{beta}].  This is our main focus because, otherwise, the description in terms of the cubic nilpotent superfield $\Phi$ becomes invalid eventually.  The inflation scale and the supersymmetry breaking scale at the vacuum will be generically the same order of magnitude.  However, it is also possible to realize a hierarchy between them as shown below and in appendix~\ref{sec:model}. 
\item The inflaton supermultiplet restores the supersymmetry at the vacuum, $\langle D_\Phi W \rangle = 0$.  In this case, eq.~\eqref{beta} becomes singular because of the vanishing $U$ in the denominator.  The use of the cubic nilpotent superfield $\Phi$ is allowed only temporarily, and the sinflaton should be restored into the theory eventually.  Moreover, the low-energy supersymmetry breaking field contributes to the Goldstino via eq.~\eqref{goldstino}, so the identity of the longitudinal gravitino is time dependent and the inflatino reappears in the physical spectrum.  Despite these complications, a potential advantage of this possibility is that it would be easier to realize a hierarchy between the inflation scale and the low-energy supersymmetry breaking scale. 
\end{enumerate}
In the remainder of the paper, we discuss the first possibility.  

 \subsection*{Example model}

Let us discuss a concrete model.
We turn off the linear coefficient $c$ in $K$ and consider the flat K\"ahler limit of the $\alpha$-scale supergravity model~\cite{Roest:2015qya}
\begin{align}
W(\Phi) =& W_0 \left( e^{\sqrt{3}\Phi} - e^{-\sqrt{3}\Phi} F\left( e^{-\frac{2\Phi} {\sqrt{3 \alpha}}} \right) \right),
\end{align}
where $\alpha > 0$ is a real positive parameter, $W_0 > 0$ is the overall normalization of the superpotential, and $F(x) = \sum_{n=0} f_n x^n$ is a real holomorphic function, i.e., all $f_n$'s are real. 
The origin of the inflaton field is a matter of convention, so we assume the minimum of the potential is located at the origin of $\Phi$.  We impose the following three conditions: (1) the self-consistency condition about the assumption of the VEV just mentioned $V'(0) = 0$, (2) the small positive cosmological constant $V(0) = \Lambda$, and (3) the tunable supersymmetry breaking parameter (controlling the ratio between the gravitino mass and the inflaton mass) $F(1) = 1 - \delta$.  Note that $\delta = 0$ (as in the original model~\cite{Roest:2015qya}) under conditions (1) and (2) with $\Lambda \to 0$ implies a supersymmetric vacuum, so we assume $\delta \neq 0$ for the application to the constrained superfield $\Phi$. 
Note that this construction with $\Lambda > 0$ does not contradict the no-go statements in Refs.~\cite{Kallosh:2014oja, Linde:2014ela} since the inflaton breaks supersymmetry also at the vacuum in our setup.

 For illustration, we truncate the function $F$ at the third order and impose the three conditions.  The $f_0$, $f_1$, and $f_2$ can be solved in terms of $\alpha$, $\delta$, and $\Lambda$.
The gravitino mass parameter $m_{3/2} = e^{K/2}W$ at the vacuum is given by $W_0 \delta$.  The inflaton mass squared at the vacuum is given by
\begin{align}
m_\phi^2 =& \frac{4  (2 + 9\sqrt{\alpha} + 9\alpha)}{3\alpha \delta} \left( (2 - \delta) m_{3/2} \sqrt{3 (3 m_{3/2}^2+  \Lambda)} -  \delta (3 m_{3/2}^2 + \Lambda) \right), 
\end{align}
where we discarded the solution of $f_0, f_1$, and $f_2$ that is always tachyonic.  
Note that the sign of $m_{3/2}=W_0 \delta$ correlates with that of $\delta$.  
For a fixed $W_0$, the supersymmetry restoration limit $\delta \to 0$ (which we do not take exactly) implies $|m_{3/2}|, m_\phi \to 0$.  For a finite $\delta$, the hierarchy $|m_{3/2}|/m_\phi$ is controlled by $\delta$.  For a more general function $F$ without truncation, the inflaton mass would be adjustable also for small $|m_{3/2}|$.  The gravitino can be heavier than the inflaton for $\delta \simeq 1$ and for $\delta \lesssim - 8\alpha^{-1} (2 + 9 \sqrt{\alpha} + 9 \alpha) $. 
As usual, the supersymmetry breaking scale at the vacuum is given by $|D_\Phi W|^2 = 3 m_{3/2}^2 + \Lambda \simeq 3 m_{3/2}^2$. 
The full expression of the scalar potential is rather complicated, but its form is similar to that of the $\alpha$-attractor models~\cite{Kallosh:2013yoa, Galante:2014ifa, Carrasco:2015pla}. It predicts the spectral index $n_\text{s} = 1 - 2 / N$ and the tensor-to-scalar ratio $r = 12 \alpha / N^2$, where $N$ is the $e$-folding number between the horizon exit of the pivot-scale mode and the end of inflation.  These are consistent with the cosmological microwave background data~\cite{Akrami:2018odb}.  For $|\delta| \ll 1$, the bottom of the inflaton potential is approximated by a quartic term as indicated by a suppressed inflaton mass in this limit.  This deformed $\alpha$-attractor-like potential is also consistent with the observational data (see, e.g., Refs.~\cite{Kallosh:2013yoa, Ketov:2016gej}).
As already mentioned, more general models can be embedded in the present setup, i.e., supergravity with a single constrained superfield by using the model-construction method of Ref.~\cite{Ferrara:2016vzg}.  

An alternative way to construct a supergravity inflation model is to use the framework of Refs.~\cite{Ketov:2014qha, Ketov:2014hya}.  An example inflation model is presented and analyzed in appendix~\ref{sec:model}.  However, a technical complication arises after inflation due to complex parameters in $K$ and $W$: the gravitino mass parameter develops the inflaton-dependent phase, which works similar to the chemical potential and leads to a modification of the dispersion relation~\cite{Kahn:2015mla}.  This requires a generalization of the previous analyses of the gravitino Lagrangian and the equations of motion which assumed the reality of the parameters and the inflaton trajectory.  Since this complication is not tied to the presence of constrained superfields and they can also arise in the linearly realized supergravity models, we study these cases in appendix~\ref{sec:complex}.

\section{On the gravitino problems \label{sec:gravitino}}

Although the numbers of  degrees of freedom are the same in the theory of a single chiral superfield $\Phi$ with $(\Phi - \bar{\Phi})^3=0$ and in the theory of two orthogonal nilpotent chiral superfields~\cite{Komargodski:2009rz, Kahn:2015mla, Ferrara:2015tyn, Carrasco:2015iij} $X$ and $T$ with $X^2 = X(T-\bar{T}) = 0$ as discussed in Ref.~\cite{Aldabergenov:2021obf}, it is important how these degrees of freedom are distributed in superfields. 
This is a crucial difference in terms of the property of the longitudinal mode of the gravitino after inflation.  In the former theory, the inflatino and the Goldstino are identical and it is just eaten by the gravitino, whereas in the latter theory, the inflatino in $T$ is removed by the constraint and the Goldstino in $X$ is eaten by the gravitino. 

To discuss this issue further, let us consider the Lagrangian density of the canonically normalized longitudinal gravitino $\psi^\ell$~\cite{Kallosh:1999jj, Giudice:1999yt, Giudice:1999am, Kallosh:2000ve, Nilles:2001ry, Nilles:2001fg, Ema:2016oxl} in the unitary gauge~\eqref{unitary_gauge} and with the conformal metric $\mathrm{d}s^2 = a(\eta)^2 ( - \mathrm{d} \eta^2 + \mathrm{d} \vec x^2)$,
\begin{align}
\mathcal{L} = & -  \frac{1}{2}\overline{\psi^\ell} \left(  \gamma^0 \partial_0 - \widehat{c}_{3/2} \left( \vec \gamma \cdot \vec \nabla \right) + a \widehat{m}_{3/2} \right) \psi^\ell,
\end{align}
where the sound-speed parameter $\widehat{c}_{3/2}$ and the effective mass parameter $\widehat{m}_{3/2}$ are 
\begin{align}
\widehat{c}_{3/2} \equiv & \frac{p_\text{SB} - \gamma^0 p_W}{\rho_\text{SB}} , \label{c3/2hat} \\
\widehat{m}_{3/2} \equiv & \frac{3 H p_W + m_{3/2} \left( \rho_\mathrm{SB} + 3 p_\mathrm{SB} \right) }{2 \rho_\text{SB}}.  \label{m3/2hat}
\end{align}
The parameter $m_{3/2} = e^{K/2}W = W$ is the mass of the transverse gravitino.\footnote{
In this section, we consider the case in which $m_{3/2}$ is real as in the example model in section~\ref{sec:inflation}.  The physical mass is its absolute value $|m_{3/2}|$. The general case is studied in appendix~\ref{sec:complex}.
}  $\rho_\mathrm{SB} \equiv \rho + 3 m_{3/2}^2 M_\text{P}^2$ and $p_\mathrm{SB} = p - 3 m_{3/2}^2 M_\text{P}^2$ are the supersymmetry breaking contributions to the energy density $\rho$ and the pressure $p$, respectively, and $p_W$ is defined as $p_W \equiv 2 \dot{m_{3/2}} M_\text{P}^2$ with a dot denoting the time derivative. 

In the original version of the minimal supergravity inflation~\cite{Ferrara:2015tyn, Carrasco:2015iij}, the orthogonal nilpotent superfields $X$ and $T$ were utilized.  It turns out that the sound-speed parameter changes~\cite{Lebedev:1989rz, Kratzert:2003cr, Hoyos:2012dh, Benakli:2014bpa, Benakli:2013ava, Kahn:2015mla, Ferrara:2015tyn} 
 significantly when the vacuum mass $m_{3/2}$ is much smaller than the inflationary energy scale $H$, and this leads to a pathologically catastrophic gravitino production~\cite{Hasegawa:2017hgd, Kolb:2021xfn, Kolb:2021nob}.  By this, we mean that gravitino modes with any wavenumber $k$ below the cutoff of the theory, whatever the cutoff is, are produced (if we neglect the backreaction).  This implies either a substantial gravitino production or breakdown of the effective theory utilizing the non-linear supersymmetry.  

On the other hand, it is also known that (the absolute value of) the sound-speed parameter is unity ($| \widehat{c}_{3/2}|^2 = (p_\text{SB}^2 + p_W^2) / \rho_\text{SB}^2 = 1$) when there is only a single relevant chiral superfield~\cite{Kallosh:1999jj, Giudice:1999yt, Giudice:1999am, Kallosh:2000ve, Nilles:2001ry, Nilles:2001fg, Ema:2016oxl}.  Since the model in section~\ref{sec:inflation} involves only the single chiral superfield $\Phi$, it is always possible to rotate the gravitino field by a $\gamma^0$-dependent phase to let $\widehat{c}_{3/2} = 1$. 
In fact, it is easy to explicitly check $p_\text{SB}^2 + p_W^2 = \rho_\text{SB}^2$ in our setup.   Even without the field redefinition, we can see that the propagation speed is given by $|\widehat{c}_{3/2}|^2 = 1$ as shown in Appendix~\ref{sec:dispersion}. Therefore, there is no catastrophic gravitino production due to the sound-speed change in our model.

Nevertheless, we need to consider the standard gravitino problem~\cite{Pagels:1981ke, Weinberg:1982zq, Khlopov:1984pf, Ellis:1984eq} which is not related to the sound speed.  This is because the gravitino may be produced in the standard inflaton decay channel or through the scattering processes in the thermal bath and because the abundance of gravitinos is tightly constrained by cosmological observations such as dark matter abundance~\cite{Moroi:1993mb} and the light-element abundance~\cite{Moroi:1995fs,Jedamzik:2004er, Kawasaki:2004yh, Kawasaki:2004qu, Jedamzik:2006xz, Kawasaki:2008qe}.
In our model, the inflaton breaks supersymmetry also at the vacuum and it is the main contribution to the supersymmetry breaking, so the inflaton decay rate into a pair of the longitudinal gravitinos is sizable~\cite{Endo:2006zj, Nakamura:2006uc, Kawasaki:2006gs, Asaka:2006bv, Kawasaki:2006hm, Endo:2007sz}, 
\begin{align}
\Gamma (\phi \to \psi_{3/2} \psi_{3/2}) \simeq \frac{m_\phi^5}{96 \pi m_{3/2}^2},
\end{align}
if it is kinematically allowed.  In the standard scenario, this leads to too large effects on the big bang nucleosynthesis or the overproduction of the lightest supersymmetric particles, which exceed the observed abundance of dark matter.  To avoid exclusion, one typically has to assume $R$-parity violation or thermal inflation~\cite{Lyth:1995ka}. 

Even if the above decay is kinematically forbidden, the thermal production of gravitinos~\cite{Bolz:2000fu,Pradler:2006qh,Rychkov:2007uq, Ellis:2015jpg, Garcia:2017tuj, Eberl:2020fml} should also be taken into account.  If the gravitino masses are much larger than the maximum cosmic temperature after inflation, their abundance is suppressed by the Boltzmann factor.  For the heavy gravitino phenomenology, see Refs.~\cite{Benakli:2017whb, Dudas:2017rpa, Dudas:2017kfz}.

\section{Discussions \label{sec:discussion}}

We have shown that the recently proposed constrained superfield~\cite{Aldabergenov:2021obf}, which is nilpotent with degree three, can be used to describe inflation in supergravity with minimal degrees of freedom, namely the inflaton, graviton, and massive gravitino, without encountering the problem of the catastrophic production of slow gravitinos~\cite{Hasegawa:2017hgd, Kolb:2021xfn, Kolb:2021nob}.  This resurrects the idea of minimal supergravity inflation~\cite{Ferrara:2015tyn, Carrasco:2015iij} (see also Refs.~\cite{Delacretaz:2016nhw, Kahn:2015mla, DallAgata:2015zxp}) in a different setup from the original construction with the orthogonal nilpotent superfields.
However, this does not automatically mean the viability of the model, and the model is constrained by the standard gravitino problem.  The final abundance of the gravitinos depends primarily on the mass spectrum of the model and also on the thermal history of the universe.  

Since there is more than one way to describe inflationary physics with minimal degrees of freedom in supergravity, it is natural to ask whether there are more.  It is important to note that the multiple descriptions are not necessarily equivalent.  Indeed, the physics of the single-superfield theory with the nilpotency condition of degree three discussed in this paper is free of the sound-speed issue and different from the physics of the orthogonal nilpotent superfields because of the different couplings between the gravitino and the matter fields.  If we start from a low-energy EFT, there can be multiple ways of UV completion.  On the other hand, there should be a unique low-energy EFT of a given UV theory.  Which low-energy theory is the correct low-energy limit is thus a UV-dependent question.  In light of the catastrophic gravitino problem~\cite{Hasegawa:2017hgd, Kolb:2021xfn, Kolb:2021nob}, it is desirable to further clarify the relations between non-linearly realized supersymmetric theories and the linearly realized ones.  In this direction, see, e.g., Refs.~\cite{Kallosh:2016hcm, Ferrara:2016een, Dudas:2021njv}.

We have exploited the unitary gauge to simplify our calculations and discussions.  If we do not take the unitary gauge, the sinflaton component will be expressed by  a complicated expression generalizing eqs.~\eqref{constraint_solution} and \eqref{beta}.  In particular, the expression would involve supercovariant derivatives of the inflatino/Goldstino $\chi$.  Even if we cannot eliminate the derivative $\partial_\mu \chi$ or the gravitino $\psi_\mu$ appearing in the supercovariant derivative simultaneously with the non-derivative $\chi$ in the unitary gauge, they appear together in the expression of $\Sigma$, so the whole combinations vanish in the unitary gauge.  This situation is similar to the case of the orthogonal nilpotent superfields~\cite{Ferrara:2015tyn}.  An analysis without specifying a gauge may give us more insights into the structure of the theory or the complementary understanding of the off-shell interactions, but we do not go further since we are not interested in gauge-dependent issues. 

 In our analysis, we have assumed that the supersymmetry is solely broken by the inflaton superfield, and the other sectors potentially existing in the full theory have been neglected.  Let us discuss how our discussion is affected when we remove this assumption.  
First, if the supersymmetry breaking contribution from the other sectors 
  is relatively small and parametrized by $\epsilon \ll 1$, 
  the Goldstino has a small component suppressed by $\epsilon$ other than the inflatino.  The unitary gauge $\upsilon = 0$ does not completely eliminate the inflatino $\chi$, so the solution of the constraint $(\Phi \pm \bar{\Phi})^3 = 0$ is affected by a correction suppressed by $\epsilon$.  This implies that the correction to the sound speed, if any, is suppressed by $\epsilon$, so the conclusion is not affected. 

Second, there is a reason to expect that the sound speed does not change even for unsuppressed $\epsilon$.  A general analysis of the gravitino system taking into account mixing with other fermions shows that the sound-speed matrix (the multifield generalization of the sound-speed parameter) can be always diagonalized into the unit matrix in the absence of constrained superfields~\cite{Nilles:2001fg, Ema:2016oxl, Roberts:2021plm}.  The sound-speed change in the theory with orthogonal nilpotent superfields can be understood as a brute-force intervention to the diagonalization process by removing a dynamical fermion by the constraint.  This point was also emphasized in Ref.~\cite{Dudas:2021njv}.  
Since the inflatino in the cubic nilpotent superfield is not removed by the constraint but just absorbed (partly, in the multi-superfield case) by the massive gravitino, we expect that the same diagonalization process is possible. Strictly speaking, the fermionic part of the Lagrangian is modified in the presence of a constrained superfield and multiple fermion species since eq.~\eqref{unitary_gauge} becomes no longer valid.  Thus, whether this intuition also applies to the case with a constrained superfield is  a nontrivial issue and should be explicitly checked elsewhere. 

\section*{Acknowledgments}
The author thanks Pablo Soler for discussions and comments, Kazunori Nakayama for discussions, Ryo Yokokura for introducing Ref.~\cite{Kugo:1982cu}, and Kyohei Mukaida, Gary Shiu, and Yusuke Yamada for discussions on a related topic.  He also benefited from a useful note by Yohei Ema~\cite{Ema_note}.
This work was supported by IBS under the project code, IBS-R018-D1.


\appendix

\section{Inflation model with complex parameters \label{sec:model}}

In this appendix, we consider an inflation model with complex parameters since the presence of complex parameters or, more precisely, the non vanishing value of $G_i \partial_\mu \phi^i - G_{\bar{i}} \partial_\mu \bar{\phi}^{\bar{i}}$ where $G\equiv K + \ln |W|^2$ affects the analysis of the gravitino equation of motion.  The latter is studied in appendix~\ref{sec:complex} while we summarize the mass spectrum, supersymmetry breaking, and dark energy for an example model in this section.

The model we consider is essentially one of the models in Refs.~\cite{Ketov:2014qha, Terada:2015sna}.
We consider the K\"ahler potential~\eqref{model_Kahler}, which leads to the general expression of the scalar potential in eq.~\eqref{V_general} under the constraint~\eqref{inflaton_constraint}. 
The superpotential is
\begin{align}
W = \mu \left( b - e^{-\sqrt{2} a \Phi} \right), \label{model_superpotential}
\end{align}
where $\mu$ is the overall coefficient which can be taken as real without loss of generality, $a$ is a real parameter (which should not be confused with the scale factor), and $b$ is a complex parameter. 

 The scalar potential for the canonically normalized inflaton $\phi = \text{Re}\, \Phi / \sqrt{2}$ is
\begin{align}
V = \mu^2 \left( (c^2 - 3) \left( b_\text{R} - e^{-a \phi} \right)^2 + \left( c b_\text{I} - \sqrt{2} a e^{-a\phi}\right)^2 - 3 b_\text{I}^2 \right),
\end{align}
where we decomposed $b= b_\text{R} + i b_\text{I}$ into its real and imaginary parts. This is written as a sum of a constant term, a term proportional to $e^{-a \phi}$, and a term proportional to $e^{-2 a \phi}$, so the potential can be regarded as a Starobinsky-like potential~\cite{Starobinsky:1980te, Bezrukov:2007ep} plus a cosmological constant. Parametrizing $a = \sqrt{2/(3 \alpha )}$, it is similar to the $\alpha$-attractor models~\cite{Kallosh:2013yoa, Galante:2014ifa, Carrasco:2015pla}. 
Similar to the model studied in the main text, it is consistent with the observational data.

The VEV of the inflaton is given by
\begin{align}
 e^{-a \phi}  = \frac{(c^2- 3) b_\text{R} - \sqrt{2}a c b_\text{I} }{c^2 + 2 a^2 -3},
\end{align}
which must be positive.
The inflaton mass squared is
\begin{align}
m_\phi^2 = \frac{2 a^2 \mu^2 \left( (c^2 -3) b_\text{R} - \sqrt{2} a c b_\text{I} \right)^2}{c^2 + 2 a^2 -3}.
\end{align}
This is to be compared with the gravitino mass squared,
\begin{align}
|m_{3/2}|^2 = \mu^2 \left|  \frac{a \left( 2 a b_\text{R} + \sqrt{2} c b_\text{I} \right)}{c^2 + 2 a^2 -3} + i b_\text{I} \right|^2 .
\end{align}
The cosmological constant is 
\begin{align}
\Lambda = \mu^2 \left( (\sqrt{2} a b_\text{R} + c b_\text{I})^2 (c^2 -3) - 3 b_\text{I}^2 (c^2 + 2 a^2 - 3) \right).
\end{align}
Note that this construction with $\Lambda > 0$ does not contradict the no-go statements in Refs.~\cite{Kallosh:2014oja, Linde:2014ela} since the inflaton breaks supersymmetry also at the vacuum in our setup.

Now, let us require that the potential can be written as 
\begin{align}
V = \frac{m_\phi^2}{2 a^2} (1 - e^{-a \phi})^2 + \Lambda,
\end{align}
so that the VEV of $\phi$ vanishes, where the inflaton mass is now $m_\phi^2 = 2 a^2 \mu^2 (c^2 + 2 a^2 -3)$. This is a requirement on $b_\text{R}$ and $b_\text{I}$, and there are two solutions for $c^2 > 3$ that lead to a positive potential.  Depending on the solutions of $b$, the gravitino mass is given by
\begin{align}
\frac{|m_{3/2}|^2}{\mu^2} = & \frac{1}{(c^2 -3)(9+2c^2(a^2-3)+c^4)} \left( 4a^4 c^2 + (c^2-3)^2 \Lambda + 2 a^2 (-9 + c^4 + c^2 \Lambda) \phantom{\sqrt{\Lambda}}  \right. \nonumber \\
& \qquad \qquad \left. \pm 2 \sqrt{2a^2c^2(c^2-3)(12a^4 + (c^2-3)^2 \Lambda + 2 a^2 (-9+c^2 (3 + \Lambda)))} \right).
\end{align}
The supersymmetry breaking scale $D_\Phi W$ is given by $|D_\Phi W|^2 = 3 |m_{3/2}|^2 + \Lambda \simeq 3 |m_{3/2}|^2$. 
The ratio $|m_{3/2}| / m_\phi$ scales as $1/(a c)$ in the large $a$ limit and $1/c^2$ in the large $c$ limit.

In summary, it is possible to realize a tiny positive cosmological constant $\Lambda$ by tuning the complex parameter $b$.  In addition, we may consider the following limits.
\begin{enumerate}
\item For $0 < c^2 -3 \lesssim \mathcal{O}(1)$, depending on the value of $a$, the gravitino mass can be arranged so that the inflaton decay into gravitinos is kinematically forbidden.  The gravitino yield will be significantly reduced.  The supersymmetry breaking scale is comparable with or higher than the inflation scale. 
\item In the large $a$ or large $c$ limits, the gravitino can be parametrically much lighter than the inflaton so that the supersymmetry breaking scale at the vacuum is hierarchically smaller than the inflation scale.  A drawback of this is the unsuppressed perturbative decay of the inflaton into a pair of gravitinos.
\end{enumerate}

\section{Gravitino Lagrangian with complex parameters \label{sec:complex}}

We follow Refs.~\cite{Kallosh:2000ve, Nilles:2001fg, Ema:2016oxl} to study the gravitino Lagrangian.  In the analyses of these references, the reality conditions are imposed at some point such that the auxiliary vector field of the (old-minimal) supergravity multiplet vanishes and the gravitino mass parameter is real. (See, however, Ref.~\cite{Roberts:2021plm} which takes into account the phase of the gravitino mass.)  For the setup of appendix~\ref{sec:model}, we need to break these assumptions, so we generalize the relevant part of the analysis in Ref.~\cite{Ema:2016oxl}.  We explicitly write $\Mpl$ in this appendix.  The Dirac gamma matrix $\gamma^\mu$ is defined in the Minkowski spacetime, which satisfies $\{ \gamma^\mu , \gamma^\nu \} = 2 \eta^{\mu\nu}$.

We work up to the quadratic order in gravitinos and take the unitary gauge $\upsilon = 0$ [see eqs.~\eqref{goldstino} and \eqref{unitary_gauge}].
After solving the constraint equations, i.e., the components of equations of motion without time derivatives, the spatial components of the gravitino field $\vec \psi$ can be decomposed into its transverse mode $\vec \psiT$ and its longitudinal mode $\psiL \equiv \vec \gamma \cdot \vec \psi$ as follows:
\begin{align}
\vec \psi = \vec \psiT + \frac{1}{k^2} \left( \vec k \left( \vec \gamma \cdot \vec k \right) + \frac{ia}{2} \left( 3 \vec k - \vec \gamma \left( \vec \gamma \cdot \vec k \right) \right) \left( \bm{m}_{3/2}^\dag - H \gamma^0 \right) \right) \psiL,
\end{align}
 in the Friedmann-Lema\^itre-Robertson-Walker universe with the conformal metric $\mathrm{d}s^2 = a(\eta)^2 (-\mathrm{d}\eta^2 + \mathrm{d}\vec{x}^2 )$. The mass parameter $\bm{m}_{3/2}$ is defined as $\bm{m}_{3/2} \equiv  (m_{3/2} P_\text{R} + m_{3/2}^* P_\text{L} )$, where $m_{3/2} \equiv e^{K/2 \Mpl^2} W/ \Mpl^2$ is complex, in general, and $P_\text{L} = (1+\gamma_*)/2$ and $P_\text{R}=(1 - \gamma_*)/2$.  The gravitino Lagrangian density is 
\begin{align}
	 \mathcal{L}_{3/2}
	=&
	 \left( \mathcal{L}_{t} + \mathcal{L}_{\ell}+\mathcal{L}_{\text{mix}} \right) .
\end{align}
The parts related to $\psiT$, $\psiL$, and the mixing term are given by
\begin{align}
	&e^{-1}\mathcal{L}_{t}
	=
	- \frac{1}{2a^3}  \overline{\vec \psiT} \left [ \slashed{D} + a \bm{m}_{3/2} \right ] \vec \psiT_{i}, \\
	&e^{-1}\mathcal{L}_{\ell}
	=
	-\frac{\rho_\mathrm{SB}}{4ak^2\Mpl^2}\overline{\psiL}\left[
	\gamma^{0}\partial_{0} + \left(i\vec{\gamma}\cdot\vec{k}\right)\widehat{c}_{3/2}
	-\frac{3a}{2}\left(\bm{m}^\dagger_{3/2} + H\gamma^{0}\right)\widehat{c}_{3/2} 
	- \frac{1}{2}a\bm{m}^\dagger_{3/2} + i \gamma^{0}\gamma_* A_0
	\right]\psiL, \\
	&e^{-1}\mathcal{L}_{\text{mix}}=
	\frac{\sqrt{2}}{a^2\Mpl}\overline{\psiL}\gamma^{0}   g_{i \bar{j}} \left( \partial_{0}\bar{\phi}^{\bar{j}}\chi^{i}_{\text{L}} + \partial_{0}\phi^{i}\chi^{\bar{j}}_{\text{R}} \right), 
\end{align}
where $e = \sqrt{-g}$ is the measure. 
The covariant derivative on the gravitino is $D_\mu \psi = (\partial_\mu + (3/2) a H \gamma^0  - i A_\mu \gamma_* ) \psi$, where $\gamma_* \equiv i \gamma_0 \gamma_1 \gamma_2 \gamma_3$.  
Note that the fermionic combination appearing in the mixing term vanishes in the unitary gauge in the single-superfield case of our scenario, $\mathcal{L}_\text{mix} = 0$. 
In the above expressions, $A_{\mu}$ is the auxiliary gauge field of $\text{U(1)}_R$ symmetry (vector auxiliary field of the old-minimal supergravity), 
\begin{align}
	A_\mu = & \frac{i}{4\Mpl^2}\left( K_i \partial_\mu \phi^i - K_{\bar{i}} \partial_\mu \bar{\phi}^{\bar{i}}\right).  
\end{align}
In the shift-symmetric case with the linear term~\eqref{model_Kahler}, we have 
$A_0 =  \sqrt{2} c \phi' / 4 \Mpl$ where a prime denotes the conformal time derivative. 
Another important parameter in $\mathcal{L}_{\ell}$ is the sound-speed parameter $\widehat{c}_{3/2}$, 
\begin{align}
\widehat{c}_{3/2}  \equiv  \frac{p_{\text{SB}}-2  \gamma^0 \Mpl^2 a D_0 \bm{m}_{3/2}^{\dag}}{\rho_{\text{SB}}} \equiv \frac{p_{\text{SB}}-\gamma^0 \bm{p}_W^{\dag}}{\rho_{\text{SB}}} ,
\end{align}
where $\bm{p}_W^{\dag}\equiv 2 \Mpl^2 a D_{0} \bm{m}_{3/2}^{\dag}$ and  $D_\mu \bm{m}_{3/2} \equiv (\partial_\mu +2 i \gamma_* A_\mu) \bm{m}_{3/2} $.  
We here record several formulas involving $\widehat{c}_{3/2}$, 
\begin{align}
\widehat{c}_{3/2}^{\dag} =&  \frac{1}{\rho_{\text{SB}}}  \left(p_{\text{SB}} + 2  \Mpl^2 a D_0 \bm{m}_{3/2}  \gamma^0 \right) , \\
\overline{\widehat{c}_{3/2}} \equiv \beta \widehat{c}_{3/2}^{\dag} \beta = & \frac{1}{\rho_{\text{SB}}}  \left(p_{\text{SB}} + 2 \gamma^0  \Mpl^2 a D_0 \bm{m}_{3/2}  \right) , \\
(\vec \gamma \cdot \vec k) \overline{\widehat{c}_{3/2}}  = & \widehat{c}_{3/2} (\vec \gamma \cdot \vec k),
\end{align}
where $\beta \equiv i \gamma^0$, and we have used $\bm{m}_{3/2} \overleftarrow{D}_{0} = D_0 \bm{m}_{3/2}$ because $\gamma_* \bm{m}_{3/2} = \bm{m}_{3/2} \gamma_*$.

Now that we have introduced various variables, let us discuss the transverse gravitino Lagrangian as a warm-up before we discuss the longitudinal one. 
By conformal rescaling, we can canonically normalize the transverse mode as $\vecpsiT_{\text{c}} \equiv \sqrt{a}\vecpsiT$,
\begin{align}
	\mathcal L_t = -\frac{1}{2}\overline{\vecpsiT_{\text{c}}} \left[ \gamma^0 \partial_0 + i \left( \vec\gamma \cdot\vec k \right) + a \bm{m}_{3/2} - i \gamma^0 \gamma_{*} A_0  \right] \vecpsiT_{\text{c}}.
	\label{Ltrans}
\end{align}
The gradient term has the same weight as the time-derivative term, so the sound speed is unity for the transverse mode. 
Note that the gravitino mass has a nontrivial phase.  The role of this phase was studied in Ref.~\cite{Kahn:2015mla}, and the shift of the wavenumber in the dispersion relation was found.  
As noted in the reference, one can optionally erase this phase by the redefinition of the gravitino, which gives rise to a new term in the Lagrangian.  
Writing $\bm{m}_{3/2}= |m_{3/2}| (e^{2i\theta_{3/2}}P_R + e^{-2i\theta_{3/2}}P_L )=|m_{3/2}|e^{-2i \theta_{3/2}\gamma_*}$, we rotate $\vecpsiT_{\text{c}}= e^{i\theta_{3/2}\gamma_*}\widetilde{\vecpsiT_{\text{c}}}$ to obtain
\begin{align}
	\mathcal L_t = -\frac{1}{2}\overline{\widetilde{\vecpsiT_{\text{c}}}} \left[ \gamma^0 \partial_0 + i \left( \vec\gamma \cdot\vec k \right) + a |m_{3/2}| - i \gamma^0 \gamma_{*} \widetilde{A}_0  \right] \widetilde{\vecpsiT_{\text{c}}},
\end{align}
where
\begin{align}
\widetilde{A}_0 \equiv A_0 - \theta'_{3/2} =\frac{i}{4\Mpl^2}\left( \partial_i G \partial_\mu \phi^i - \partial_{\bar{i}} G \partial_\mu \bar{\phi}^{\bar{i}}\right)
\end{align}
where $G= K + \Mpl^2 \ln (|W|^2/\Mpl^6 )$.
In the limit $\widetilde{A}_0 = 0$, it reduces to the transverse gravitino Lagrangian in Ref.~\cite{Ema:2016oxl}.  
Let us see the effects of the new term $\widetilde{A}_0$ using the results in appendix~\ref{sec:dispersion}. 
We decompose the transverse gravitino as in eq.~\eqref{mode_decomposition} and obtain the dispersion relation~\eqref{dispersion_relation_general} for the general Lagrangian~\eqref{L_Majorana_general} with $c_0 =1$, $c_1 = c_2 = 0$, $m_0 = \text{Re}\, m_{3/2}$, $m_1 = - \text{Im}\, m_{3/2}$, and $m_2=0$,
\begin{align}
&u_{\vec k, h}^\pm{}'' - \left [ \log \left((kh  -  \widetilde{A}_0 )/\Mpl\right) \right ]' u_{\vec k, h}^\pm{}'  \nonumber \\
&+ \left ( (a |m_{3/2}|)^2 + \left(k + h \widetilde{A}_0\right)^2 \mp i (a |m_{3/2}|)' \pm i  \left [ \log \left((kh  -  \widetilde{A}_0 )/\Mpl \right) \right ]' a |m_{3/2}| \right ) u_{\vec k, h}^\pm = 0.
\end{align}
In the Minkowski spacetime limit $a=1$ with a constant $\widetilde{A}_0$, the dispersion relation reduces to $\omega^2_k = |m_{3/2}|^2 + \left( k \pm \widetilde{A}_0 \right)^2$, which coincides with the result for a Majorana spin-1/2 fermion in Ref.~\cite{Kahn:2015mla}.

Let us now move on to the longitudinal mode.  First, we canonically normalize the field
\begin{align}
	\psiLc \equiv - \frac{\sqrt{\rho_{\rm SB}} a^{3/2}}{\sqrt{2}k^2\Mpl} i\left( \vec\gamma \cdot\vec k\right)\psiL.
\end{align}
The Lagrangian becomes
\begin{align}
	\mathcal L_\ell =& - \frac{1}{2}\overline{\psiLc} \left[ 
		\gamma^0\partial_0 - i\left( \vec\gamma \cdot\vec k\right) \overline{ \widehat{c}_{3/2}}  
		+\frac{3a}{2}\left(\bm{m}_{3/2} - H\gamma^{0}\right) \overline{\widehat{c}_{3/2}} 
	+ \frac{1}{2}a \bm{m}_{3/2} - i \gamma^{0}\gamma_* \widetilde{A}_0
	\right]\psiLc. \nonumber \\
	&=  - \frac{1}{2}\overline{\psiLc} \left[ 
		\gamma^0\partial_0 - i\left( \vec\gamma \cdot\vec k\right) \overline{ \widehat{c}_{3/2}}  
		+\widehat{\bm{m}}_{3/2} - i \gamma^{0}\gamma_* \widetilde{A}_0 \left( 1  - \frac{6 |m_{3/2}|^2 \Mpl^2 }{\rho_\text{SB}} \right)
	\right]\psiLc, 
\end{align}
where 
\begin{align}
\widehat{\bm{m}}_{3/2} \equiv \frac{3 H \bm{p}_W + \bm{m}_{3/2} \left(\rho_\text{SB} + 3 p_\text{SB} \right)  }{2\rho_\text{SB}}
\end{align}
is a generalization of $\widehat{m}_{3/2}$ in eq.~\eqref{m3/2hat}.   
Using the fact that
\begin{align}
D_\mu \bm{m}_{3/2} = & e^{K/2\Mpl^2} (D_i W \partial_\mu \phi^i P_R + \bar{D}_{\bar{i}}\bar{W} \partial_\mu \bar{\phi}^{\bar{i}} P_L)/\Mpl^2, 
\end{align}
one can show
\begin{align}
\widehat{c}_{3/2}^\dag \widehat{c}_{3/2} = & \frac{p_\text{SB}^2 + 4 a^2 D_0 \bm{m}_{3/2} D_0 \bm{m}_{3/2}^\dag }{\rho_\text{SB}^2} = 1 = \widehat{c}_{3/2} \widehat{c}_{3/2}^\dag .
\end{align}
This implies that $\widehat{c}_{3/2}$ can be written as $ - \widehat{c}_{3/2} = e^{2 \gamma^0 ( \theta P_\text{R} + \theta^* P_\text{L}) } $ with $\theta$ a complex number, which also implies $-\overline{\widehat{c}_{3/2}} = e^{-2 (\theta P_\text{R} + \theta^* P_\text{L})\gamma^0 }$.  
We redefine the longitudinal mode as $\psiLc \equiv e^{-(\theta P_\text{R} + \theta^* P_\text{L})\gamma^0} \widetilde{\psiLc} \equiv U \widetilde{\psiLc}$ to diagonalize the gradient term. The new Lagrangian reads
\begin{align}
	\mathcal L_\ell =
	&=  - \frac{1}{2}\overline{\widetilde{\psiLc}} \left[ 
		\gamma^0\partial_0 + i\left( \vec\gamma \cdot\vec k\right) 
		+ \bar{U}\widehat{\bm{m}}_{3/2} U - i \gamma^{0}\gamma_* \widetilde{A}_0 \left( 1  - \frac{6 |m_{3/2}|^2 \Mpl^2 }{\rho_\text{SB}} \right) U^2
- \gamma^0 U^\dag \partial_0
	 U \right]\widetilde{\psiLc}.
\end{align}
This shows that the sound speed of the longitudinal mode is also equal to the speed of light even in the presence of complex parameters.  
The effective mass term has the structure $m_0 \mathbb{1} + m_1 \gamma_* + m_2 \gamma^0 \gamma_*$.  We study the dispersion relation of the fermion containing such terms in appendix~\ref{sec:dispersion}.  

\section{Generalized dispersion relation for a Majorana spinor \label{sec:dispersion}}
After the decomposition of the gravitino field into its transverse and longitudinal modes, each equation of motion essentially reduces to that of a spin-1/2 Majorana fermion.  In general setups, the Lagrangian contains complex parameters and the equation of motion can contain nontrivial terms such as $\gamma^0\gamma_*$-dependent mass terms. In this appendix, we study such a generalized Lagrangian for a Majorana fermion $\psi$.

We consider the Lagrangian 
\begin{align}
\mathcal{L} = & - \frac{1}{2} \overline{\psi} \left(  \gamma^0 \partial_0 + i (\vec  \gamma \cdot \vec k) (c_0 + c_1 \gamma^0  + i c_2 \gamma^0 \gamma_* ) + m_0 + i m_1 \gamma_* + i m_2 \gamma^0 \gamma_* \right)  \psi, \label{L_Majorana_general}
\end{align}
where $c_0$, $c_1$, $c_2$, $m_0$, $m_1$, and $m_2$ are real parameters that can be time dependent. 
We decompose the fermion field into spinor modes as in Ref.~\cite{Ema:2016oxl, Hasegawa:2017nks},
\begin{align}
\psi = \sum_{h = \pm 1} \int \frac{\mathrm{d}^3 k }{(2\pi)^{3/2}} e^{i \vec k \cdot \vec x} \begin{pmatrix}
u^{+}_{\vec k, h} (\eta) \\
u^{-}_{\vec k, h} (\eta) 
\end{pmatrix}
\otimes \xi_{\vec k, h} \hat{b}_{\vec k,h} + \text{H.c.}, \label{mode_decomposition}
\end{align}
where $\xi_{\vec k, h}$ satisfies $(\vec \sigma \cdot \vec k) \xi_{\vec k, h} = h k \xi_{\vec k, h}$, $\hat{b}_{\vec k, h}$ is the annihilation operator, and H.c.~denotes the Hermitian conjugate.
Combining the first-order mode equations for $u^{\pm}_{\vec k, h} (\eta)$,
\begin{align}
u^{\pm}_{\vec k, h}{}' \mp i \left( a m_0 + h c_{2} k \right) u^{\pm}_{\vec k, h} = &  \left( a(m_1 \pm i m_2 ) - i h (c_0 \pm i c_1) k  \right) u^{\mp}_{\vec k, h} , 
\end{align}
 we obtain its second-order mode equation as follows:
\begin{align}
u^{\pm}_{\vec k, h}{}'' - & \log [ kh (c_0 \pm i c_1 )   \mp i a m_1 + a m_2 ]' u^{\pm}_{\vec k, h}{}'  + \left( a^2 m_\text{eff}^2 + c_\text{eff}^2 k^2 \mp i(a m_0+ h c_2 k )' \right. \nonumber \\
& \left.   + 2h k a (\vec c \cdot \vec m) \pm i  \log [ kh (c_0 \pm i c_1 ) \mp i a m_1 + a m_2 ]' (a m_0 + h c_2 k )  \right)  u^{\pm}_{\vec k, h} = 0, \label{dispersion_relation_general}
\end{align}
where $m_\text{eff}^2 \equiv m_0^2 + m_1^2 + m_2^2$, $c_\text{eff}^2 \equiv c_0^2 + c_1^2 + c_2^2$, and $\vec c \cdot \vec m \equiv c_0 m_2 + c_1 m_1 + c_2 m_0$. 

Some comments are in order.
First of all, if we set $c_1 = c_2 = m_1 = m_2 = 0$ with $c_0 = \pm c (= \pm 1)$ and $m_0 \equiv m$, this reduces to the standard dispersion relation $(\omega_k/a)^2 = m^2 + c^2  (k / a)^2$.  More generally, the dispersion relation contains more terms and splits with signs.  
In the case of the transverse gravitino, $c_0 =1$, $c_1 = c_2 = 0$, $m_0 = \text{Re}\, m_{3/2}$, $m_1 = - \text{Im}\, m_{3/2}$, and $m_2=0$.  The longitudinal gravitino can have nonzero values for any of the six parameters.  As we saw in appendix~\ref{sec:complex}, we can set $c_0 = 1$, $c_1 = c_2 = 0$ by field redefinition. In the Minkowski limit $a=1$ and with constant parameters, time-derivative terms drop and the expression simplifies to $\omega_{k,h}^2 = ( k \vec c + h \vec m )^2$ where $\vec c = (c_0, c_1, c_2)$ and $\vec m = (m_2, m_1, m_0)$.  This reproduces a part of the results in appendix C of Ref.~\cite{Kahn:2015mla} where constant $m_2$ is introduced as a non-minimal term.  However, we do not find superluminality  due to $m_1$ or $m_2$ found in the reference provided that $c_0^2 + c_1^2 + c_2^2 \leq 1$.  When we set $c_1 = m_1 = m_2 =0$ with a time-dependent $c_0$, it reproduces the dispersion relation found in Ref.~\cite{Hasegawa:2017nks} for the case of orthogonal nilpotent superfields with real parameters.  In this case, the singularity due to the vanishing argument of the $\log$ leads to particle production.  

Intriguingly, the logarithmic-derivative part also depends on $m_1$ and $m_2$, which implies that nontrivial time dependence of $m_1$ and $m_2$ can lead to particle production.  A qualitatively different feature is that the singularity does not happen for sufficiently large $k$ modes since $k$ also appears in the $\log$.  This precludes the potential unlimited production of UV modes.  It will be interesting to study the phenomenological and cosmological applications of the particle production based on the time dependence of these non-minimal parameters $c_1$, $c_2$, $m_1$, and $m_2$, which naturally appear in the supergravity context. This has been studied in Ref.~\cite{Roberts:2021plm}, based on Refs.~\cite{Adshead:2018oaa, Adshead:2019aac}, in the two-superfield case without the matrices corresponding to $c_2$ and $m_1$ to study the fermion production during inflation.  We wish to study various applications in future work. 

\small

\bibliographystyle{utphys}
\bibliography{MSIr.bib}

\end{document}